\documentclass[
reprint,
superscriptaddress,
 amsmath,amssymb,
prb,
]{revtex4-2}

\usepackage{graphicx}% Include figure files
\usepackage{dcolumn}% Align table columns on decimal point
\usepackage{bm}% bold math
\begin{document}

\preprint{APS/123-QED}

\title{Signatures of Enhanced Superconducting Properties in Niobium Cavities}% Force line breaks with \\

\author{D. Bafia}
 \altaffiliation{dbafia@fnal.gov}
\affiliation{
 Fermi National Accelerator Laboratory, Batavia, Illinois 60510, USA}
\author{A. Grassellino}
\affiliation{
 Fermi National Accelerator Laboratory, Batavia, Illinois 60510, USA}
\author{M. Checchin}
\affiliation{
 SLAC National Accelerator Laboratory, Menlo Park, California 94025, USA}
\author{J. F. Zasadzinski}
\affiliation{
 Department of Physics, Illinois Institute of Technology, 
 Chicago, Illinois 60616, USA}
\author{A. Romanenko}
\affiliation{
 Fermi National Accelerator Laboratory, Batavia, Illinois 60510, USA}

\date{\today}

\begin{abstract}

Superconducting radio-frequency (SRF) niobium cavities are critical for modern particle accelerators, as well as for advancing superconducting quantum systems and enabling ultra-sensitive searches for new physics. In this work, we report a systematic observation of an anomalous frequency dip in Nb cavities, which occurs at temperatures just below the critical temperature ($T_\mathrm{c}$), indicative of enhanced superconducting properties at $T \ll T_c$. The magnitude of this dip is strongly correlated with the RF surface resistance, impurity distribution near the surface, and $T_\mathrm{c}$. Additionally, we report measurements of the coherence peak in the AC conductivity of two Nb SRF cavities processed using distinct methods. By comparing recent theories developed to model this experimental data, we show that the frequency dip feature, larger coherence peak height, and reduction in the temperature-dependent surface resistance with RF current occur at minimal but finite levels of disorder.
\end{abstract}

\maketitle

\section{Introduction}

Superconducting radio frequency (SRF) niobium cavities are record-high quality factor $Q_0>10^{10}-10^{11}$ man-made resonators that serve as the primary accelerating structures in modern particle accelerators~\cite{Padamsee_Ann_Rev_Nucl_2014}, longest coherence microwave quantum superconducting systems~\cite{Rom20}, and an ultrasensitive platform for dark sector and dark matter candidate searches~\cite{Janish_PRD_2019, Bogorad_PRL_2019, Asher_JHEP_2020, Gao_arxiv_2020}. These ultra-high quality factors are enabled by the low microwave surface resistance of superconducting Nb, given by  $R_s = G/Q_0$, where $G$ is a geometric factor. Typically, $R_s$ increases with the magnitude of the peak surface magnetic field $B_p$ due to increased screening currents. Extensive research over the past decades has allowed the discovery and mitigation of several such field-dependent physical mechanisms in various regimes~\cite{Padamsee_Ann_Rev_Nucl_2014}. This most recently includes proximity-coupled niobium nanohydrides~\cite{Rom13} and their mitigation by oxygen inward diffusion~\cite{Romanenko_SRF2019_THP014} and the reduction of two-level system-driven dissipation by oxide removal~\cite{Rom17, Rom20,Bafia_PRApplied_2024}. 

This steady improvement has revealed new SRF phenomena that remain poorly understood, most notably the factor of four increase in $Q_0$ and the field-dependent \emph{increase} in the quality factor of nitrogen-doped~\cite{Grass13} cavities, which is commonly referred to as the positive slope Q. The unusual effect stems from an anomalous decrease in the temperature-dependent component of the surface resistance $R_T$ with increasing field~\cite{Grass13}. In essence, it is as if superconducting pairing strengthens with an increased RF field and current. Material studies have highlighted its dependence on the concentration and distribution of nitrogen in the vicinity of the surface. Possible mechanisms include a field-stimulated smearing of the BCS density of states \cite{Garfunkel_PR_1968,Gurevich_PRL_2014}, anomalous skin effects \cite{XIAO_2013}, and the field-driven nonequilibrium redistribution of thermally excited quasiparticles to higher energy levels \cite{Martinello_PRL_2018}. Still, a systematic study of the effect of doping on quasiparticle lifetimes and density of states as well as strong coupling pairing effects in SRF cavities has not been reported and is critical for the understanding of the effect. 

\begin{figure}[h!!]
\includegraphics[width=6.9cm]{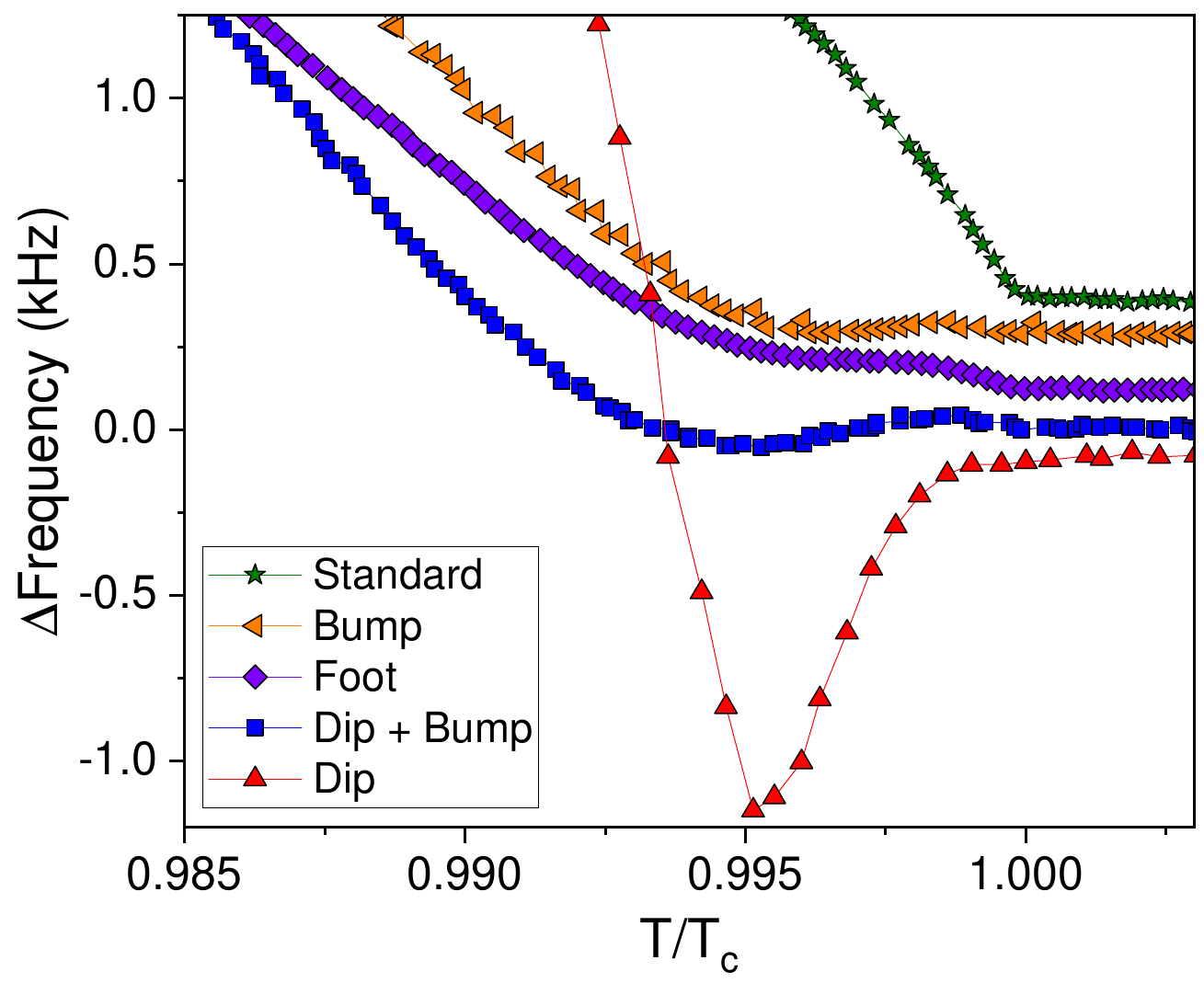}
    \caption{Five observed resonant frequency variations of 1.3~GHz elliptical niobium SRF cavities near $T_c$. Each frequency curve has been offset for clarity.}
    \label{fig:Zoology}
\end{figure}

While many studies have focused on the quality factor and surface resistance, here we report that the cavity resonant frequency provides another valuable avenue for gaining insight into the underlying physical phenomena, as its changes are directly connected to the properties of the superconducting condensate. We find the existence of anomalous frequency features just below the critical transition temperature $T_c$ of the bulk superconducting Nb. Fig. \ref{fig:Zoology} shows the five resonant frequency variations just below the critical transition temperature ($T_c$) in Nb SRF cavities. All five features have been observed at low fields ($<$1~mT) while warming and cooling through $T_c$, which is expected as the superconducting phase transition is of the second order.

The AC complex conductivity enables further insight on quasiparticles and superconducting carriers. For frequencies far below the superconducting gap ($\hbar \omega \ll \Delta$), $\sigma_1$ exhibits the coherence peak, a nonmonotonic dependence of the quasiparticle conductivity with temperature that arises due to coherence factors in BCS theory \cite{BCS_theory} and is analogous to the Hebel-Slichter peak in nuclear spin relaxation \cite{Hebel_PR_1959}. This peak was originally demonstrated in niobium by Klein \textit{et al.} \cite{Klein_PRB_1994} at 60~GHz, a frequency significantly higher than those typically used in modern RF accelerators. We show that the height of the coherence peak also correlates with the frequency dip in Fig.\ref{fig:Zoology}.

In this article, we present a systematic study on a large number ($>$40) of bulk niobium SRF cavities revealing the anomalous decrease (dip) in cavity resonant frequency at temperatures just below $T_\mathrm{c}$ in nitrogen-doped cavities. In addition, we report that the other four (non-dip) characteristic temperature dependencies uniquely correspond to other explored surface treatments, and thus correlate strongly with cavity performance in $R_T$ with the stored field. The measured coherence peak in the complex conductivity of bulk niobium SRF cavities is fitted with a phenomenological model incorporating a pair-breaking term in the BCS \cite{BCS_theory} density of states. We comment on different models proposed by other groups which help provide insight on the origins of these various phenomena. Fits with theories suggest that the coherence peak, along with the dip phenomenon, high $Q_0$, and $R_T$ decrease with field, are signatures of improved superconducting properties and occur when extrinsic pair-breaking scattering and pair-conserving scattering from nonmagnetic impurities are minimal but finite.  

\section{Experimental Procedure}

A large set of bulk niobium SRF cavities of TESLA~\cite{Aune00} elliptical shape were employed for our studies. The resonant frequencies of these cavities varied from 650~MHz to 3.9~GHz, which allowed probing the low frequency limit of Nb ($\hbar \omega \ll \Delta$). All cavities went through a bulk electropolishing (EP) of 120~$\mu$m material removal and 800$^\circ$C hydrogen degassing step \cite{Padamsee98} followed by various surface treatments. These treatments included additional $\sim$40~$\mu$m EP, nitrogen infusion~\cite{Grassellino_SUST_2017}, 120$^\circ$C baking for 48 hours \cite{Hasan_book2}, 75/120$^\circ$C baking \cite{Grassellino_75120C_arXiv_2018, Bafia75/120C} wherein cavities were first baked at 75$^\circ$C before the standard 120$^\circ$C bake, and nitrogen doping~\cite{Grass13} with varying dopant concentrations. Dopant concentration is varied by altering the durations of baking in N atmosphere and in vacuum, which has the shorthand notation of minutes in N/minutes in vacuum (i.e., 2/6) and/or the amount of removal via EP. 

All cavities were first cooled down to 2~K where standard measurements~\cite{Melnychuk_RSI_2014} of $Q_0$ as a function of the peak surface magnetic field $B_\mathrm{peak}$ have been performed; most cavities were tested again after cooling further down to $T<$~1.5~K. To extract quasiparticle driven surface resistance $R_T$ from the measured $Q_0$, we used the methods laid out in \cite{Martinello_PRL_2018}. Specifically, we subtracted off the residual resistance ($R_{\text{res}}=G/Q_0(<1.5 \text{K})$) contribution from the total $R_s$ at 2~K, so that $R_{T}=R_s(2 \text{K})-R_{\text{res}}$.

Measurements of the resonant frequency $f_0$ with temperature were made as follows. Cavities were equipped with resistance temperature detectors (RTDs), placed in a helium dewar, and pre-cooled to 4.2 K. We used heaters located at the bottom of dewar to boil off the liquid helium and facilitate warming. This procedure allowed for a warming rate of $<0.1$~K/min, ensuring thermalization of the cavity, as confirmed by the RTD readings. Resonant frequency was recorded with a vector network analyzer (VNA). Measurements persisted through the niobium superconducting transition temperature at $\sim$9.2~K. After measurements, the data was corrected to account for variations in dewar gas pressure.

For several $f_0(T)$ datasets, an effective mean free path $\ell$ within the cavity surface layer has been obtained by converting the measured frequency shift with temperature near $T_\mathrm{c}$ into a shift in the magnetic field penetration depth $\lambda(T)$ using Slater's theorem~\cite{Slater46} and fitting based on the modified Halbritter routine~\cite{Hal70}, following the same technique outlined in~\cite{Martinello_APL_2016}. 

To calculate the AC conductivity, we used the method utilized by Trunin \textit{et al.} \cite{Trunin97}. To extract the surface resistance, we measured $Q_0$ from 1.5~K up to 10~K using a combination of standard power balance measurements in continuous wave and decay modes along with VNA measurements and converted to surface resistance $R_s$. The surface reactance was measured utilizing the above discussed frequency vs temperature measurements from 6~K to 10~K and extrapolating to lower temperatures by means of fitting with the Halbritter routine~\cite{Hal70} and using $X_s (T) = -2 G\Delta f_0 (T)/f_0 + X_n$. We obtained the additive constant using the normal conducting surface resistance measured just above the transition temperature $R_n$(10 K). In the local limit, valid for dirty Nb, $R_n(\text{10 K})=X_n( \text{10 K})$. When the local limit is not applicable, as is true for high purity niobium \cite{Hasan_book2}, the correct relationship between $R_s$ and $X_s$ is obtained using the universal impedance curves plotted against the dimensionless parameter $\alpha$ calculated by Reuter and Sondheimer in the microwave region \cite{RS48}. The parameter $\alpha$ is proportional to $\ell^3$/$\rho \ell$, where $\ell$ is the mean free path and $\rho \ell$ is a temperature independent material constant, which is 6~$\times$~10\textsuperscript{-16}~$\Omega$m\textsuperscript{2} for Nb \cite{Hasan_book2}. Finally, the complex conductivity is calculated as 
\begin{equation}
    \sigma=\sigma_1+i \sigma_2=\omega \mu _0 \bigg( \frac{2R_sX_s}{(R_s^2+X_s^2)^2}+ i\frac{X_s^2-R_s^2}{(R_s^2+X_s^2)^2} \bigg).
    \label{eq:sig}
\end{equation}

\section{Results}

\subsection{Effect of Impurity Structure on Frequency Variations Near $T_c$}

Overall, we have studied the frequency response of 41 cavities with various impurity structures in the RF layer. Of these 41, we find a clear dip in the resonant frequency just below the $T_\mathrm{c}$ in 22 N-doped cavities, similar to what is observed in Fig.~\ref{fig:Zoology}. For other treatments, we observed the four other characteristic behaviors.

\begin{table}[h!] 
\centering
      \begin{tabular}{c|c |  c |  c |  c |  c }
           \hline         & N-doped & N-Infused & 75/120$^{\circ}$C & 120$^{\circ}$C & EP\\
          \hline
          \hline
          Dip & 22 & &1\footnote{$\Delta f_{dip}$ $<$ 50 Hz as compared to $\sim$1-2~kHz in doped cavities, likely an artifact of the pressure correction}  & \\
          Foot & & 1 & 5 &  \\
          Bump & & & 1& \\
          Dip+Bump & & 2 & & & 1\\
          Standard & & 1 & 3 & 2 & 2\\
          \hline
      \end{tabular}
      \caption{Occurrence of features in $f_0(T)$ data near $T_c$ for five typical SRF cavity surface treatments.} 
      \label{tab:featStats}
  \end{table}

\subsection{Effect of MFP on Frequency Dip Feature}

\begin{figure}[h!]
    \centering
    \includegraphics[width=7.3cm]{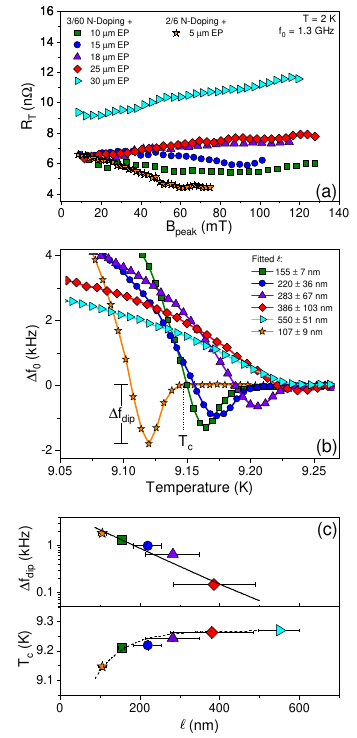}
    \caption{(a) Temperature dependent surface resistance at 2~K with peak magnetic field and (b) resonant frequency response with temperature of a single 1.3 GHz Nb cavity after sequential removal of the surface post nitrogen doping. The legend in (b) presents the fitted $\ell$ in the RF layer after each step. Resonant frequency shift is reported relative to the normal conducting value near 10 K. The parameter $\Delta f_{dip}$ is called the dip magnitude. (c) Dip magnitude and $T_c$ are plotted against the fitted $\ell$. Horizontal error bars come from fitting; vertical error bars are smaller than the data points. Solid and dashed lines show $e^{-\ell}$ and $-e^{-\ell}$ relationships, respectively.}
    \label{fig:SeqRem}
\end{figure}

To investigate the dependence of the dip on the nitrogen concentration, one cavity subjected to 3/60 nitrogen doping underwent several sequential material removal steps of the RF surface via EP, with RF measurements performed after each step. The material removal allowed to gradually decrease the concentration of N present in the RF layer, resulting in an increase in $\ell$. After the combined 30~$\mu$m material removal, the cavity was bulk electropolished and re-processed with the 2/6 N-doping surface treatment. The $R_T (B_\mathrm{peak})$ and $f_0(T)$ results along with the corresponding material removal amounts and extracted $\ell$ values are presented in Fig.~\ref{fig:SeqRem}. 

We observe that dilute and uniform concentrations of nitrogen (short $\ell$) produce a characteristic decrease in $R_T$ at higher $B_\mathrm{peak}$, a weak suppression in $T_c$, and an anomalous frequency dip just before $T_c$. As nitrogen concentration gradually decreases with more material removal, these behaviors are gradually diminished. The $R_T$ behavior with field approaches that of a typical EP cavity. Meanwhile, the dip magnitude diminishes as the transition temperature assumes the clean niobium value. Subsequent bulk electropolishing and 2/6 doping, which is known to produce a larger nitrogen concentration than 3/60 doping, shows a return of the dip feature with a larger magnitude and lower $T_\mathrm{c}$.

Fig.~\ref{fig:SeqRem}(c) plots the dip magnitude and transition temperature against $\ell$. The dip magnitude decreases strongly (close to exponentially) while the transition temperature increases with increasing $\ell$ values. The $T_c$ varies by $\sim$120~mK throughout the course of the study, which is in good agreement with previous work on the effects of nitrogen and oxygen on niobium $T_\mathrm{c}$~\cite{Desorbo63}. Possible origins for this $T_\mathrm{c}$ suppression in the presence of N interstitial may stem from lattice parameter expansion \cite{Desorbo63} or Fermi surface anisotropy \cite{Zarea_FrontierPhysics_2023}. Our findings clearly show that a frequency dip and corresponding $T_c$ suppression occur in the presence of uniform and dilute concentrations of N impurities and that impurity concentration serves as a critical parameter in determining the extent of the phenomena.

\subsection{Correlating $Q_0$ with $\Delta f_{dip}$}

\begin{figure}[h!]
    \centering
    \includegraphics[width=7.2cm]{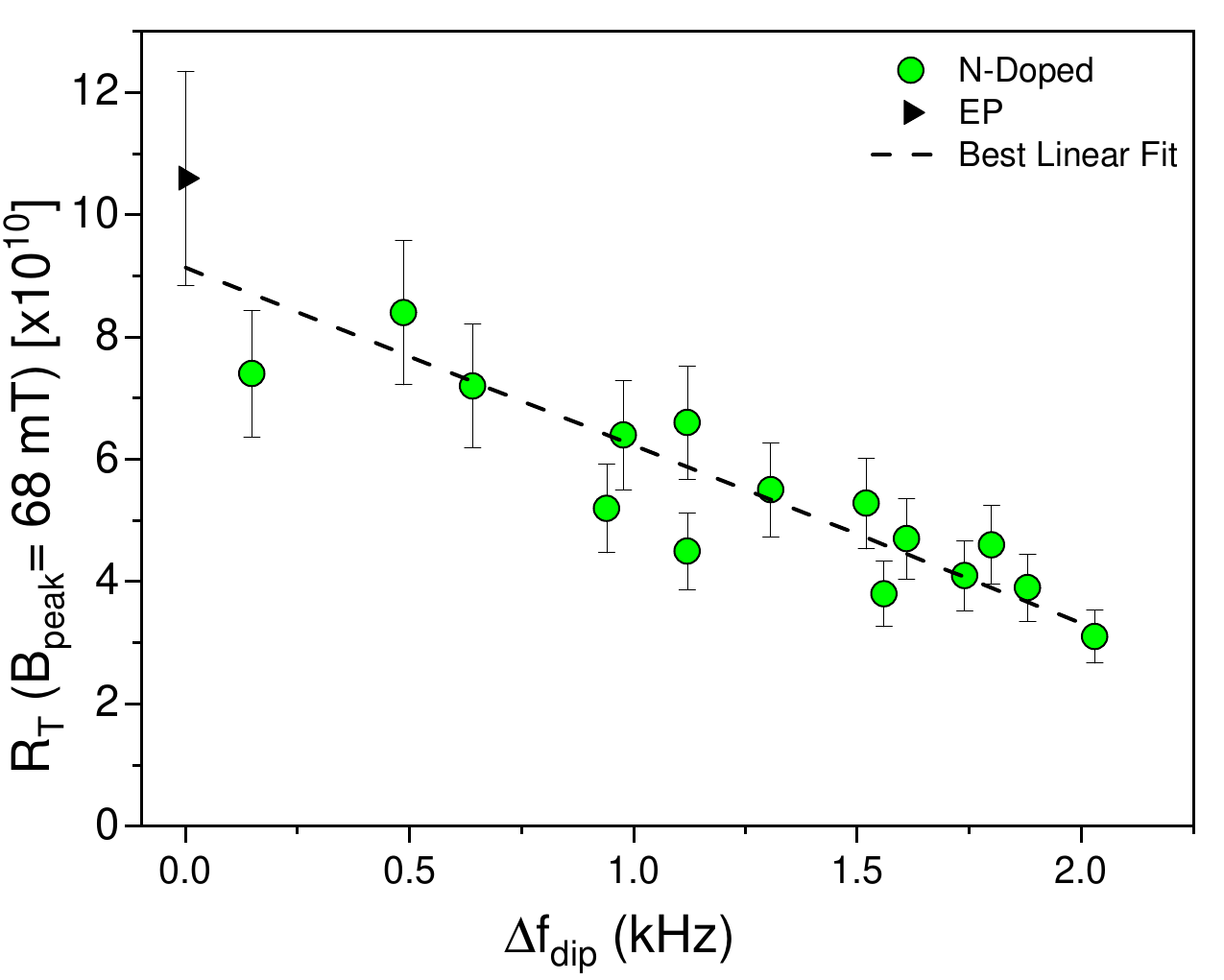}
    \caption{Temperature dependent surface resistance measured at 68~mT and a temperature of 2~K plotted against the dip magnitude for variously nitrogen doped cavities. Shown for reference are results obtained from a cavity post EP.}
    \label{fig:QvsFDip}
\end{figure}

The evolution observed in Fig.~\ref{fig:SeqRem} suggests correlation between the frequency dip and $R_T$; to investigate this, we plotted $R_T$ measured at 68~mT and 2~K against the dip magnitude for an ensemble of variously nitrogen doped cavities in Fig.~\ref{fig:QvsFDip}. Indeed, we observe a linear trend, suggesting a potential relationship between the positive Q-slope and frequency dip phenomena. We note that the trend shown in Fig.~\ref{fig:QvsFDip} is relevant for the accelerator community as it facilitates a method for quickly estimating the temperature-dependent component of the surface resistance of N-doped cavities at high fields from low field frequency domain measurements, which may be practical when high power infrastructure is not readily available.

\subsection{Effect of Resonant Frequency on Dip}

\begin{figure}[h!]
    \centering
    \includegraphics[width=7.8cm]{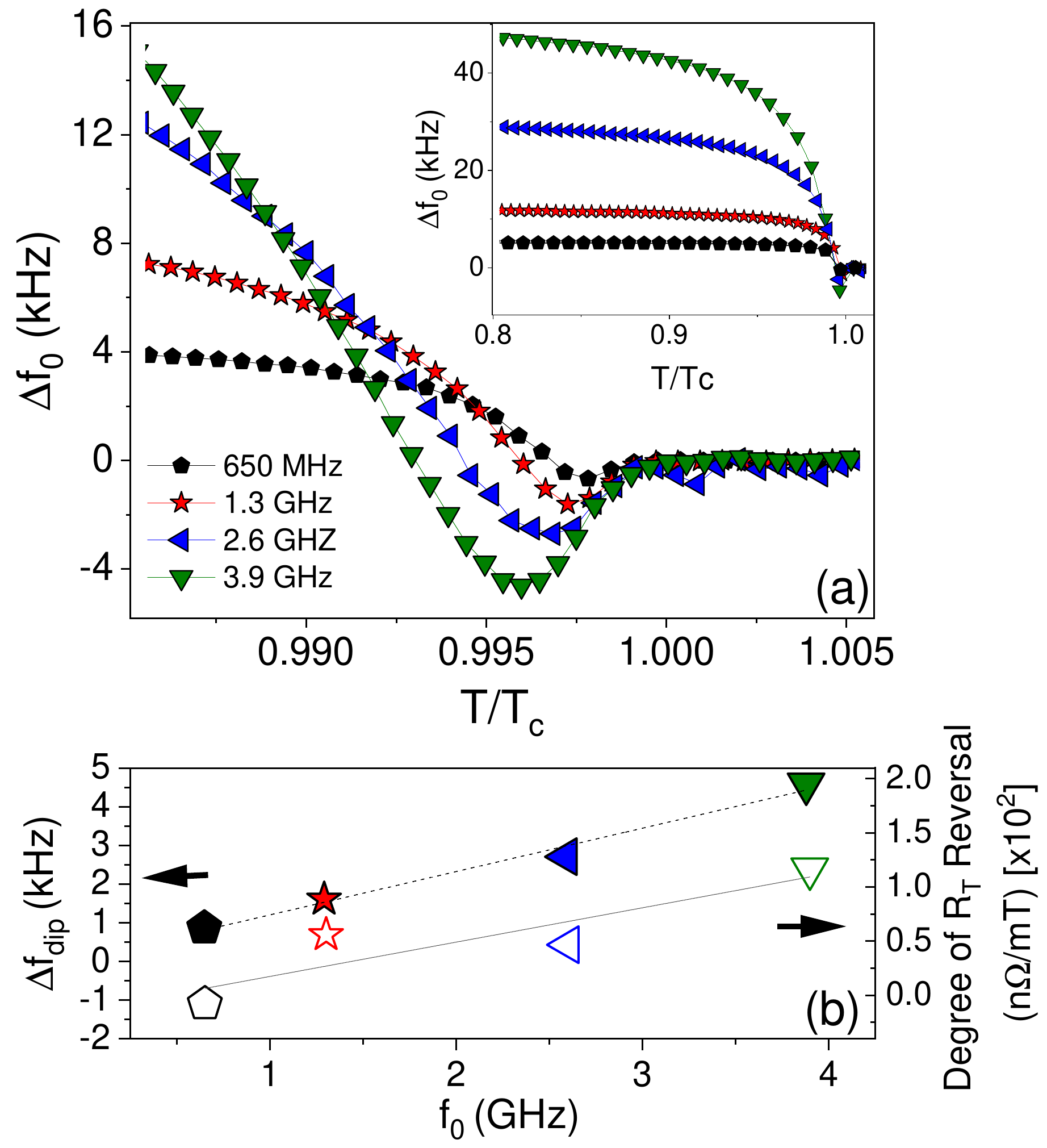}
    \caption{(a) Frequency response with temperature of four cavities, each with different resonant frequency, processed to yield an identical nitrogen concentration in the RF layer. Inset shows the response over a larger range of temperatures. (b) Left-hand axis shows the dip magnitude plotted against the fundamental resonant frequency for each cavity. The right-hand axis plots the degree of reversal of the temperature dependent component of the surface resistance for each cavity, as defined in the text. Dashed and solid lines show linear fits.} 
    \label{fig:varFreq}
\end{figure}

To study the effect of resonant frequency on the dip, four niobium cavities with $f_0$=650~MHz, 1.3~GHz, 2.6~GHz, and 3.9~GHz were subjected to the same 2/6 N-doping surface treatment, producing an identical impurity structure in each cavity. The results are shown in Fig.~\ref{fig:varFreq}(a). We observe that resonant frequency is linearly related with the dip magnitude, as shown in  Fig.~\ref{fig:varFreq}(b). Furthermore, the degree of $R_T$ reversal, defined here as the negative of the slope from 15~mT to 64~mT of the $R_T(B_{\text{peak}},T=\text{2~K})$ curve normalized to the value at 15~mT, also exhibits a linear trend with the resonant frequency. A similar dependence of the $R_T$ reversal is shown in \cite{Martinello_PRL_2018} where it is hypothesized that such behavior may be driven by non-equilibrium superconductivity. 

\subsection{Measurement of AC Complex Conductivity}

\begin{figure}[h!]
    \centering
    \includegraphics[width=6.8cm]{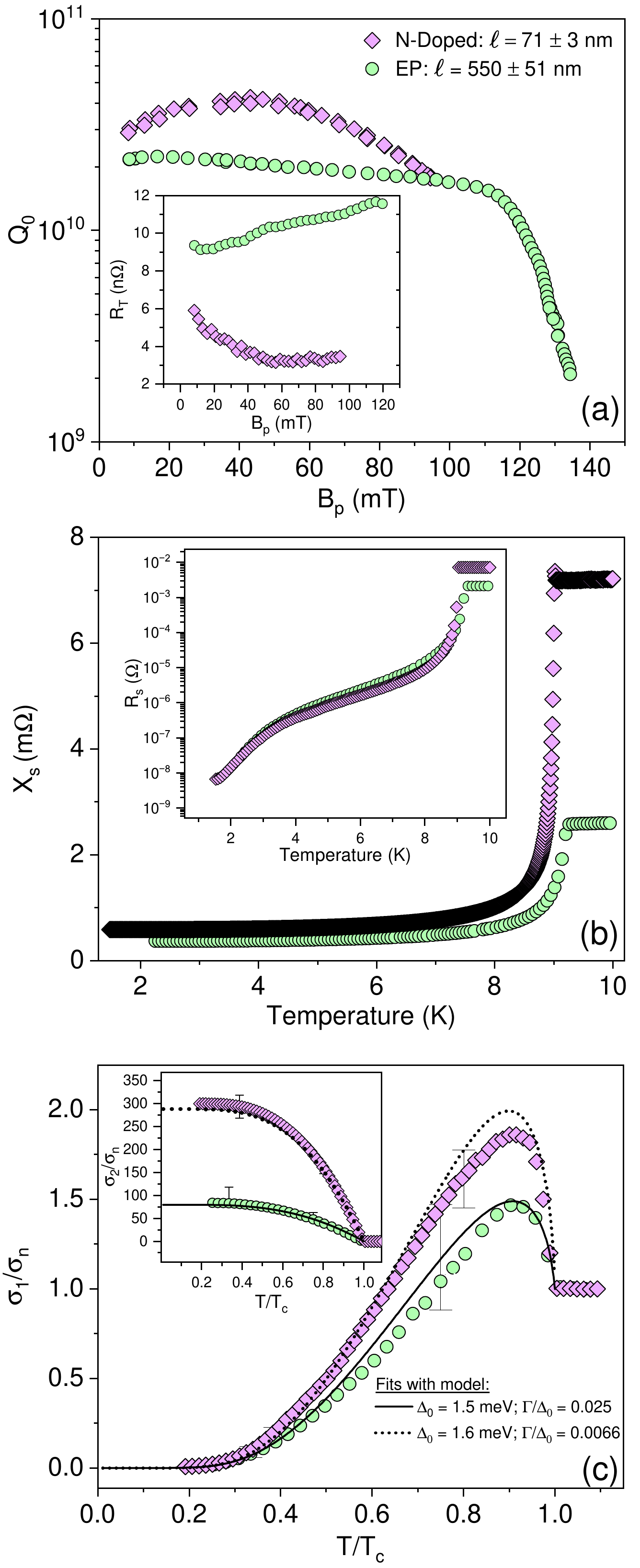}
    \caption{(a) Quality factor vs peak magnetic field of two 1.3~GHz cavities subjected to EP or N-doping. Inset shows the temperature dependent surface resistance at 2~K as a function of peak magnetic field. (b) Surface reactance of the EP and N-doped cavities as a function of temperature; inset depicts the surface resistance as a function of temperature. (c) Real ($\sigma_1$) and imaginary ($\sigma_2$) components of the AC conductivity normalized to the normal conducting value  $\sigma_n$. Error bars show the uncertainty in the calculated values (see text for details). Dashed and solid lines show fits.}
    \label{fig:cond}
\end{figure}

To gain insight on the conditions under which different frequency features near $T_c$ occur, we measured the AC complex conductivity of two niobium cavities: i) the cavity presented in Fig.~\ref{fig:SeqRem}(a) with $\ell=550\pm51$~nm which showed the standard feature near $T_c$ (which we call the EP cavity) and ii) a cavity post nitrogen doping which exhibited the prominent dip and $\ell=71\pm3$~nm. We note that this latter mean free path is in agreement with the value obtained by extrapolating Fig.~\ref{fig:SeqRem}(c) down to $T_c=$~9.04~K, yielding approximately 70~nm. Fig.~\ref{fig:cond}(a) and its inset show the expected $Q_0(B_p)$ and $R_T(B_p)$ behaviors characteristic of cavities subjected to these surface treatment. Fig.~\ref{fig:cond}(b) presents the real and the imaginary parts of the measure surface impedance of the two cavities.

To confirm the validity of our measured reactance, we calculated the experimental penetration depth at $T=$~0~K via $X_{s,0} = \omega \mu_0 \lambda_{0,\text{exp}}$ using the Halbritter routine to extrapolate data to 0~K and compared with the effective penetration depth $\lambda_{\text{eff}} = \lambda _L(1+\xi_0/\ell)^{1/2}$. For Nb, $\lambda _{L}$ = 39 nm and $\xi_0$ = 38 nm \cite{Maxfield_Phys_Rev65}. The experimental and calculated results shown in Table \ref{tab:CompExpTh} are in good agreement.

\begin{table}[h!]
\centering
      \begin{tabular}{c|c |  c |  c |  c }
           \hline Treatment & $X_{s,0}$ [$\Omega$] &$\ell$ [nm] & $\lambda_{0,\text{exp}}$ [nm] &  $\lambda_{\text{eff}}$ [nm]\\
          \hline
          \hline
          EP & 3.70 $\times$ 10 \textsuperscript{-4} & 550 $\pm$ 51 & 36 $\pm$ 4 & 40 $\pm$ 0.1\\
          N-Doped & 5.88 $\times$ 10 \textsuperscript{-4} & 71 $\pm$ 3 & 57 $\pm$ 6 & 48 $\pm$ 0.4\\
          \hline
      \end{tabular}
      \caption{Comparison of experimentally and theoretically obtained values for the penetration depth at $T$ = 0 K.} 
      \label{tab:CompExpTh}
  \end{table}

The calculated AC complex conductivities of the N-doped and EP cavities are presented in Fig.~\ref{fig:cond}(c). The error bars indicate measurement uncertainty. For the N-doped cavity, this uncertainty is primarily due to the measurement process itself, whereas for the EP cavity, the dominant uncertainty comes from the fitted value of $\ell$. We note that as the technique used to measure the surface impedance sums over the inner RF surface, these curves represent the average cavity response. 

The non-monotonic dependence in the real part of the complex conductivity in BCS superconductors, called the coherence peak, arises due to the singularity in the superconducting density of states at the gap edge \cite{BCS_theory, Mattis_Surf_Res_1958}. According to Dynes \textit{et al.} \cite{Dyn78}, inelastic scattering smears the singularity and introduces subgap quasiparticle states. By incorporating the Dynes smearing parameter $\Gamma$ into a modified version of the Mattis-Bardeen conductivity, it is possible to obtain a phenolomenological measure of the pair-breaking processes present in a system by fitting the coherence peak using

\begin{equation}
\begin{split}
    \frac{\sigma _1 }{\sigma _n} =\frac{2}{\hbar \omega} \int_{\Delta}^{\infty}\frac{
    [f(E)-f(E+\hbar \omega ) ]g(E,\Gamma)}{\sqrt{(E+i\Gamma)^2-\Delta^2}}dE
    \\
    +
    \frac{1}{\hbar \omega}\int_{\Delta-\hbar \omega}^{-\Delta}
    \frac{[1-2f(E+\hbar \omega) ]g(E,\Gamma)}{\sqrt{(E+i\Gamma)^2-\Delta^2}}dE,
   \label{eq:DynesMBCond1}
\end{split}
\end{equation}
\begin{equation}
    \frac{\sigma _2 }{\sigma _n} =\frac{2}{\hbar \omega} \int_{\Delta- \hbar \omega,-\Delta}^{\Delta}
    \frac{[1-2f(E+\hbar \omega )]g(E,\Gamma)}{\sqrt{(\Delta^2 - (E+i\Gamma)^2}}dE,
    \label{eq:DynesMBCond2}
\end{equation}
\begin{equation}
    g(E,\Gamma) = \frac{(E+i\Gamma)((E+i\Gamma)+ \hbar \omega) + \Delta ^2}
    {\sqrt{((E+i\Gamma)+\hbar \omega)^2-\Delta^2}
    }.  
    \label{eq:g}
\end{equation}

From Fig. \ref{fig:cond}(c), we observe that both the EP and N-doped cavities exhibit the so-called coherence peak, with the maxima at $\sim$0.9~$T/T_c$ differing in amplitude. This suggests a variation in the quasiparticle behavior within the RF layer due to distinct impurity distributions. The stark difference observed in $\sigma_2/\sigma_n$ between the two cavities highlights a variance in the superconducting condensate behavior. 

The dashed and solid curves in Fig.~\ref{fig:cond}(c) are calculated using Eq.~\ref{eq:DynesMBCond1}-\ref{eq:g}. We find that both $\sigma_1/\sigma_n$ and $\sigma_2/\sigma_n$ of the EP cavity are best modelled with an average superconducting gap $\Delta_0=1.5$~meV and an inelastic scattering parameter $\Gamma/\Delta_0$~=~0.025. The N-doped cavity is instead better fitted with a larger average superconducting gap $\Delta_0=1.6$~meV and a lower level of pair-breaking ($\Gamma/\Delta_0$~=~0.0066). Note that a larger coherence peak occurs for the cavity with greater elastic scattering (shorter $\ell$) due to the presence of N interstitial. According to Anderson's theorem \cite{Anderson_PRL_1959}, elastic scattering off nonmagnetic impurities has no effect on superconductivity. As a result, while N impurities increase scattering, they likely mitigate depairing processes which results in a larger coherence peak. Notably, when the fit parameters from the N-doped data are applied to Eq.~\ref{eq:DynesMBCond1}-\ref{eq:g} at 60~GHz, the calculated $\sigma_1/\sigma_n$ results roughly match Klein's local-limit measurements \cite{Klein_PRB_1994}.

\section{Discussion} 

The frequency phenomena and trends observed in this work have stimulated the development of new theoretical approaches which seek to provide insight into their microscopic origins. Ueki \textit{et al.} provided the first explanation of the anomalous near $T_c$ frequency phenomena \cite{Ueki_2022} by calculating the complex surface impedance via Slater's approach to solving Maxwell's equations for the case of a cavity while considering anisotropy in the superconducting gap \cite {Zarea_FrontierPhysics_2023} and inhomogeneous disorder within the penetration depth. By varying the level of disorder within the cavity screening region, Ueki calculated the foot, dip+bump, and dip features in the frequency response just before $T_c$. Moreover, the authors obtained the predicted cavity $Q_0$ as a function of disorder. We note that the frequency dip feature and a maximum in $Q_0$ were obtained for intermediate levels of disorder. Ueki also correctly reproduced the frequency dependence of the dip feature reported in Fig.~\ref{fig:varFreq}. Zarea \textit{et al}, instead, derived the eigenvalue equation for the fundamental TM mode of a cylindrical RF cavity which considers the current response obtained with the Keldysh formulation of the quasiclassical theory of superconductivity \cite{Zarea_FrontElectronicMaterials_2023}. The result considers the confinement of the normal conducting and superconducting currents within the RF layer and shows that the anomalous frequency dip is a product of the competition between the normal metal skin depth and the London penetration depth. An increase in impurity concentration not only affects the penetration depth but also the classification of superconducting niobium from a weak type-I to a weak type-II superconductor \cite{Prozorov_PRB_2022}. As a result, the presence of interstitial nitrogen can influence this competition between length scales. We note that Zarea also found that the frequency dip and a maximum in cavity $Q_0$ occur at intermediate levels of disorder.

Taken together, the theories by Ueki and Zarea indeed support the observation that a dilute concentration of nitrogen in  Nb, which provides an intermediate level of disorder, is responsible for the anomalous frequency dip just before $T_c$. Moreover, they report that greater levels of disorder yield a larger dip magnitude and an increase in $Q_0$ toward its peak value, giving one model for the origins of the linear correlation observed in Fig.~\ref{fig:QvsFDip}.

The results presented in Fig.~\ref{fig:cond} have also stimulated theoretical work by other groups which further improves our understanding of the mechanisms behind the improved performance of niobium SRF cavities in the presence of various near-surface impurity profiles. Herman and Hlubina \cite{Herman_PRB_2021} used the Nam approach \cite{Nam_PhysRev_1967} to calculate the optical response of a superconductor in the limit where energy tends to zero and with pair-conserving and pair-breaking elastic-scattering processes built into the energy and gap functions. The result describes the effect of subgap quasiparticle states on the optical conductivity and utilizes the concept of ``Dynes superconductors." The authors present excellent fits to the real and imaginary parts of the AC complex conductivity of both cavities presented in Fig.~\ref{fig:cond} and find that pair-breaking processes dominate the height of the coherence peak in $\sigma_1/\sigma_n$. Herman and Hlubina also highlight that the ratio between pair-breaking and pair-conserving scattering rates is key in determining the coherence peak extent and show that cleaner superconductors counterintuitively yield lower peak heights. Kubo, instead, used the Dynes $\Gamma$ in the Eilenberger formalism of the BCS theory to calculate several superconductor quantities in the presence of varying levels of nonmagnetic impurity scattering rates as a function of temperature and energy \cite{Kubo_PRApplied_2022}. These calculations also well model the results presented in Fig.~\ref{fig:cond} and, similar to Herman and Hlubina, show that the ratio between pair-breaking and nonmagnetic impurity scattering rates is key in determining the height of the coherence peak. Moreover, Kubo finds that there exists a minimum in the surface resistance at finite levels of pair-breaking and pair-conserving scattering. We note that the phenomenologically derived Eq. \ref{eq:DynesMBCond1}-\ref{eq:g} in the present manuscript reproduce the salient feature of the above theories that greater smearing of the density of states yields a reduction of the coherence peak height and the zero-temperature value of $\sigma_2/\sigma_n$. 

The above experimental and theoretical observation that different impurity structures in Nb yield varying RF performance due to the level of scattering within the RF surface is consistent with point contact tunneling spectroscopy (PCTS) studies by Groll \textit{et al.} on similarly treated cavity cutouts~\cite{Gro18}. The N-doped samples in those studies show more homogeneous $\Delta _0$ values and lower levels of $\Gamma/\Delta_0$ than those which come from EP cavities \cite{PCTSvsRF}. The lower level of pair-breaking scattering processes observed in Fig.~\ref{fig:cond} for the N-doped cavity, as most obviously denoted by a higher coherence peak and zero-temperature value of $\sigma_2/\sigma_n$, is likely due to the absence (or lower volume fraction) of proximity coupled nanohydrides and magnetic moments within the interface. This is consistent with the fact that N-doped cavities do not exhibit the high field $Q$ drop caused by the proximity breakdown of hydrides at high fields \cite{Grass13, Rom13}. 

Collectively, these findings show that the frequency dip is a hallmark of enhanced Nb SRF cavity performance, as it occurs at the intermediate levels of disorder required for the minimization of the pair-conserving and pair-breaking scattering processes. This, in turn, leads to a larger coherence peak in $\sigma_1/\sigma_n$ and an increase in the zero-temperature value of $\sigma_2/\sigma_n$. We thus conjecture that the pair-conserving scattering originates in interstitial impurities that have the beneficial effect of trapping hydrogen.

\section{Conclusion}

This study provides an investigation on the anomalous resonant frequency features that appear just below the critical temperature of Nb SRF cavities and their correlation with several metrics. We find that dilute and uniform concentrations of nitrogen in Nb yield the frequency dip phenomenon and show that the magnitude of this dip is correlated with enhanced superconducting properties, including an increased coherence peak height in AC conductivity and a reduction in surface resistance with increasing RF field. By comparing recently developed theories, we find that these phenomena are driven by an optimal level of impurity scattering, particularly off of nonmagnetic impurities such as nitrogen, which improves cavity performance. The findings suggest that nitrogen doping reduces surface irregularities, such as hydrides and magnetic oxides \cite{Bafia_PRApplied_2024}, which can otherwise impair superconducting performance. Our results not only contribute to the fundamental understanding of superconductivity in niobium but also provide practical guidelines for improving the performance of SRF cavities, with significant implications for accelerator technologies and quantum systems.

\section{Acknowledgments} 
The authors would like to acknowledge O. Melnychuk and D. A. Sergatskov for technical support during measurements. Work supported by the Fermi National Accelerator Laboratory, managed and operated by Fermi Research Alliance, LLC under Contract No. DE-AC02-07CH11359 with the U.S. Department of Energy.

\bibliography{Main}

\begin{thebibliography}{44}
\bibitem [{PCT()}]{PCTSvsRF}%
  \BibitemOpen
  \href@noop {} {}\bibinfo {note} {While PCTS reports higher values of $\Gamma/\Delta_0$ for both N-doped and EP samples than what is shown in Fig.~\ref{fig:cond}, this is likely due to differences in the probing depth between the two techniques, as PCTS is more surface sensitive and probes regions with more pair-breaking mechanisms.}\BibitemShut {Stop}%
\end{thebibliography}

\end{document}